\documentclass[conference,letterpaper]{IEEEtran}

\usepackage{layouts} 
\usepackage{cite}
\usepackage[T1]{fontenc}
\usepackage{amsmath,amssymb,amsfonts}
\usepackage{algorithmic}
\usepackage{threeparttable}
\usepackage{graphicx}
\usepackage{textcomp}
\usepackage{xcolor}
\usepackage{tabularx}
\def\BibTeX{{\rm B\kern-.05em{\sc i\kern-.025em b}\kern-.08em
    T\kern-.1667em\lower.7ex\hbox{E}\kern-.125emX}}

\begin{document}

\title{\emph{libtxsize}---a library for automated\\Bitcoin transaction-size estimates}

\author{\IEEEauthorblockN{Johannes Hofmann}
\IEEEauthorblockA{\textit{Dept. of Computer Science} \\
\textit{University Erlangen-Nuremberg}\\
Erlangen, Germany \\
johannes.hofmann@fau.de}
}

\maketitle


\begin{abstract}
    This paper presents \emph{libtxsize}, a library to estimate the size
    requirements of arbitrary Bitcoin transactions. To account for different
    use cases, the library provides estimates in bytes, virtual bytes, and
    weight units. In addition to all currently existing input, output, and
    witness types, the library also supports estimates for the anticipated
    Pay-to-Taproot transaction type, so that estimates can be used as input
    for models attempting to quantify the impact of Taproot on Bitcoin's
    scalability.

    \emph{libtxsize} is based on analytic models, whose credibility is
    established through first-principle analysis of transaction types as well
    as exhaustive empirical validation. Consequently, the paper can also serve
    as reference for different Bitcoin data and transaction types, their
    semantics, and their size requirements (both from an analytic and
    empirical point of view).
\end{abstract}

\section{Introduction}
\label{sec:intro}

In most instances today, transaction-size estimates are used to construct
favorable transactions that minimize transaction fees. These estimates rely on
heuristics and simple tools~\cite{web:lopp,web:optech,web:bbww}. The solutions
available so far are, however, limited in scope: all are limited to giving
estimates for transactions using only inputs of identical type; some are even
limited to transactions that use the same input and output type. These
shortcomings are addressed in this paper, which presents \emph{libtxsize}, a
library to give size, weight, and virtual size estimates for transactions with
arbitrary inputs, outputs, and witnesses.

Although \emph{libtxsize} can be used in the context of fee optimization, the
library was developed to provide estimates that can serve as inputs for
quantitative models whose projections provide objective data to assess the
impact of future improvements on Bitcoin's scalability. To this end,
\emph{libtxsize} includes support for the anticipated Pay-to-Taproot
transaction type.

This paper is structured as follows. Digital signatures, which are used by
Bitcoin to establish ownership of coins, are discussed and investigated as to
their size requirements in Section~\ref{sec:encodings}. The formats of
Bitcoin transactions and their components, such as inputs, outputs, and
witnesses, are presented in Section~\ref{sec:formats} and studied as to their
size requirements.  Section~\ref{sec:analysis} investigates the dynamic
components of inputs, outputs, and witnesses, derives analytic estimates, and
validates them using empirical data. Section~\ref{sec:libtxsize} discusses how
\textit{libtxsize} integrates the findings of the previous sections in a
bottom-up way. Finally, Section~\ref{sec:results} provides a summary and
concludes the paper.


\section{Encoding of public keys and signatures}
\label{sec:encodings}

Bitcoin uses public keys and signatures to establish ownership of funds. So
far, Bitcoin relies on the elliptic curve digital signature algorithm
(\textsc{ecdsa}) for signature verification. This, however, might change with
BIP~340, which introduces Schnorr signatures. Both methods rely on elliptic
curve cryptography (\textsc{ecc}), and use the same elliptic curve,
\emph{secp256k1}, defined in the Standards for Efficiency Cryptography
(\textsc{sec})~\cite{SEC1}. Despite these similarities, there are some
differences between \textsc{ecdsa} and Schnorr signatures concerning the
encoding of public keys and signatures in Bitcoin. These differences are
discussed in the following.

\subsection{Encoding of public keys}

\subsubsection{Encoding of \textsc{ecdsa} public keys}

In \textsc{ecc}, public keys correspond to the $x$- and $y$-coordinates of
points on an elliptic curve. The \emph{secp256k1} curve used by Bitcoin uses
32-byte numbers to represent the these coordinates.

In case of \textsc{ecdsa}, these coordinates are encoded using a standard
defined in the \textsc{sec}~\cite{SEC1}.  Originally, Bitcoin only supported the
uncompressed \textsc{sec} format. This format includes a one-byte \textsc{sec}
prefix, which specifies the type of encoding (in this case, the uncompressed
format), followed by the key's two 32-byte $x$- and $y$-coordinates. Encoding
a public key in the uncompressed \textsc{sec} format thus requires 65~bytes.

After some time, Bitcoin added support for the compressed \textsc{sec} format,
which reduces the size of the encoding significantly by taking advantage of
the fact that a public key's $y$-coordinate can be derived from its
$x$-coordinate (with the exception of its sign).\footnote{\label{fn:ecc-y2}For a given
$x$-coordinate, the corresponding $y$-coordinates are obtained by solving the
\emph{secp256k1} curve's equation $y^2=x^3+7$ for $y$. Note that $y^2$ implies
two valid solutions, $y$ and $-y$, which differ only with respect to sign.}
The compressed \textsc{sec} format thus only includes a public key's
$x$-coordinate.  The ambiguity concerning the $y$-coordinate's sign is
resolved using the \textsc{sec} prefix: there are two magic numbers for the
compressed \textsc{sec} format; one indicates a positive $y$-coordinate, the
other a negative one.  By getting rid of 32-byte $y$-coordinate, the
compressed \textsc{sec} format allows public keys to be encoded using only
33~bytes.

\begin{figure}[tb]
    \centerline{\includegraphics[width=\linewidth]{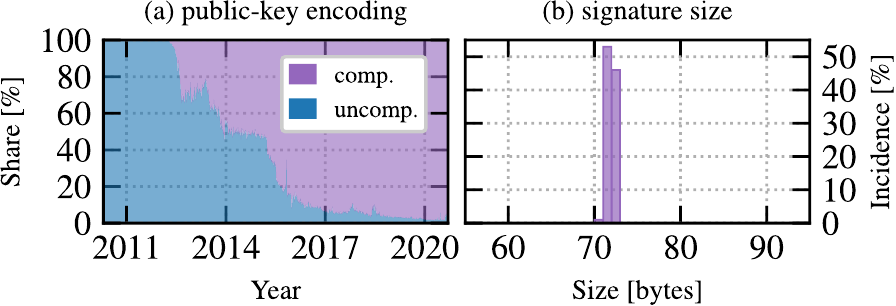}}
    \caption{(a) Public key encoding type and (b) histogram of signature sizes.}
    \label{fig:pubkey-sig}
\end{figure}

Fig.~\ref{fig:pubkey-sig}a shows the share of uncompressed and compressed
public keys in transactions over time.  The empirical data reveals a steady
decline in the use of uncompressed keys in Bitcoin over the years, a
development presumably driven by the reduction in fees for transactions using
smaller public keys. In fact, in the recent past the share of transactions
using uncompressed keys has been so low that, for practical purposes, they can
be neglected. In light of this, for the following investigation, public keys
are assumed to have a size of 33~bytes.

\subsubsection{Encoding of public keys in Schnorr signatures}

Public keys in Schnorr signatures also correspond to curve points.  For
Schnorr signatures, however, the encoding of public keys does not follow the
\textsc{sec}~\cite{SEC1} standard.  Instead, a custom encoding is used, which
includes only a point's $x$-coordinate. Ambiguity concerning the
$y$-coordinate is resolved using the coordinate's parity by defaulting to the
even $y$ value.\footnote{The modulo of curve \emph{secp256k1} is odd, so the
modulo operation maps even numbers to odd ones and vice versa.  Since the two
solutions for $y$ vary only in sign (cf.\ footnote~1), one $y$
will be even and the other odd.} The encoding of Schnorr public keys thus have
a size of 32~bytes.



\subsection{Encoding of signatures}
\label{sec:encoding:sig}

\subsubsection{Encoding of \textsc{ecdsa} signatures}

\textsc{ecdsa} signatures for the \emph{secp256k1} curve consist of two
32-byte numbers, $r$ and $s$, and Bitcoin's encoding of such signatures
follows the Distinguished Encoding Rules (\textsc{der})~\cite{DER}.

Per \textsc{der}, each of the numbers, $r$ and $s$, is prefixed by two bytes:
one to encode the data type, which in case of Bitcoin is signed integer; the
other to indicate the size of the following data---typically 32~bytes for the
values $r$ or $s$. So far, this makes for a total overhead of four bytes.

On top of that, there are two bytes at the beginning of each \textsc{der}
signature: one to indicate that the signature consists of two objects ($r$ and
$s$), and another to indicate the total size of the encoding. Finally, an
additional byte at the end of the signature indicates the signature hash type,
which is used by Bitcoin to determine which parts of a transaction to use when
creating a signature. Overall, this results in seven extra bytes to encode
the two 32-byte values $r$ and $s$.

In theory, this should result in a total signature size of 71~bytes. In
practise, however, several points must be considered. First of all, the fact
that Bitcoin uses signed integers as data type has implications on the
signature size: for, whenever $r$ or $s$ have their most significant bit
(\textsc{msb}) set, their encodings require an additional one-byte zero
padding so they are not interpreted as negative numbers. Note that values that
have their \textsc{msb} set are also referred to as ``high values,'' whereas
those that have the bit not set are referred to as ``low values.''

Originally, no restrictions applied to $r$ and $s$. Assuming uniform
distribution of values, each number should have its \textsc{msb} set half of
the time, implying a 25\% chance of no padding, a 50\% chance of padding for
exactly one of the values, and a 25\% chance of padding for both values.
In this case, signatures are expected to have an average size 72~bytes.

Today, however, several restrictions apply: Bitcoin Core 0.11.1 introduced the
``low $s$'' rule, which enforces that only transactions with low $s$ values
are relayed.\footnote{Creating a signature with one or more low values is done
by changing the random input used during signature creation for as long as is
necessary until a signature with the desired number of low values is found.}
Thus, $s$ is guaranteed to require no padding; $r$, however, still requires
padding half of the time, leading to an average signature size of 71.5~bytes.
Newer implementations voluntarily implement the ``low $r$'' optimization, and
create signatures that have a low $r$ value as well. In this case, none of the
values require padding, resulting in an average signature size of 71~bytes. 

Finally, whenever an $r$ or $s$ value has eight or more of its leading bits
set to zero, the respective value can be encoded using less than 32~bytes.
The overall signature size can thus be smaller than the previously established
averages.

To quantify the overall impact of the different factors influencing the
signature size, an analysis of empirical data is in order. The histogram in
Fig.~\ref{fig:pubkey-sig}b shows the incidence of different signature sizes
over the course of the last two years. As expected, the bulk of signatures
have a size of 71 or 72 bytes. The slight bias toward 71 bytes can be credited
to the previously discussed ``low $r$'' optimization of state-of-the-art
implementations.  Moreover, a small amount of 70-byte signatures can be
observed. These can be attributed to the previously discussed occurrence of
leading zero bits in the binary representation of $r$ and $s$ values. In fact,
there are even signatures with sizes of only 69 or 68~bytes; these, however,
occur too infrequently to be visible in the histogram. The average
signature size according to empirical data is 71.46~bytes, which is in line
with the analytically estimate of 71.5~bytes established previously.  Thus,
for the following investigation, signatures are assumed to have a size of
71.5~bytes.


\subsubsection{Encoding of Schnorr signatures}

Schnorr signatures consist of a curve point, $P$, and a 32-byte value $s$.  To
save space, the signature only contains $P$'s 32-byte $x$-coordinate from
which the corresponding $y$-coordinate can be derived (ambiguity concerning
$y$ is again addressed by implicitly using the $y$ that is even).  Unlike
\textsc{ecdsa}, Schnorr signature encoding does not follow
\textsc{der}~\cite{DER}; instead, the two 32-byte values, $r$ and $s$, are
encoded back to back without additional metadata.  The size requirement for
Schnorr signatures is thus 64~bytes.




\section{Transaction format}
\label{sec:formats}

In the following, the general formats of transactions as well as transaction
inputs, outputs, and witnesses are discussed, and the formats' implications on
transaction size are investigated.

\subsection{Transaction format}

All transactions includes the following fields: a four-byte version, two
variable-length integers\footnote{A variable-length integer is a data type
devised to minimize the size when encoding non-negative integers up to a size
of eight bytes. Its requires one byte to encode values from 0 to 252; three
bytes for up to 16-bit integers; five bytes for up to 32-bit integers; and
nine bytes for up to 64-bit integers.} (varints) to indicate the number of
inputs and outputs, and a four-byte lock-time.  Segregated Witness (SegWit)
transaction include two additional bytes: a one-byte SegWit marker to indicate
that the transaction includes witness data, and a one-byte SegWit version.
Note that no additional varint is required to indicate the number of witnesses
in the transaction, since the number of witnesses implicitly corresponds to
the number of inputs.

Transaction-size estimates thus include a fixed eight-byte (or ten-byte, in
case of SegWit) contribution; the size of the encodings of two varints; and
the sizes of the inputs and outputs (and, in case of SegWit, the witnesses).


\begin{table}[!tb]
    \caption{Locking script formats by transaction type.}
    \begin{center}
        \begin{tabularx}{\linewidth}{|l|X|}
            \hline
            \textbf{Type}       & \textbf{Locking script}   \\
            \hline
            \textsc{p2pk}       & $\sigma_p$ $p$ \textsc{op\_checksig} \\
            \textsc{p2pkh}      & \textsc{op\_dup} \textsc{op\_hash160} 20 $h$ \textsc{op\_equalverify} \textsc{op\_checksig} \\
            Bare \textsc{ms}    & \textsc{op\_}$m$ $\sigma_{p_1}$ $p_1$ \ldots\,\, $\sigma_{p_n}$ $p_n$ \textsc{op\_}$n$ \textsc{op\_checkmultisig}   \\
            Null Data           & \textsc{op\_return} $\sigma_d$ $d$                                                                             \\
            \textsc{p2sh}       & \textsc{op\_hash160} 20 $h_r$ \textsc{op\_equal}                                                 \\
            \textsc{p2wpkh}     & \textsc{op\_0} 20 $h_p$                                                                          \\
            \textsc{p2wsh}      & \textsc{op\_0} 32 $h_r$                                                                          \\
            \textsc{p2tr}       & \textsc{op\_1} 20 $h$                                                                                      \\
            \hline
        \end{tabularx}
        \label{tab:locking}
    \end{center}
\end{table}

\subsection{Transaction input format}

From a high-level viewpoint, each transaction input comprises two key pieces
of information: a reference to an unspent transaction output (\textsc{utxo})
and an unlocking script that satisfies the locking script of the referenced
\textsc{utxo}.

The \textsc{utxo} reference consists of a 32-byte transaction identifier
(\textsc{txid}) to reference a previous transaction, and a 4-byte position
number to indicate a particular output of the referenced transaction. 

Unlocking scripts have a variable size, and their contents depend on the type
of \textsc{utxo} they are trying to spend. Script formats for different
transaction types will be covered in the next section; for now, it is
sufficient to note that the script's size is encoded using a varint.

Finally, each input contains a 4-byte sequence number, which is currently used
for the replace-by-fee mechanism that allows updating a transaction's fee.

To summarize, the input size is made up a fixed component of 40~bytes,
comprising the 32-byte \textsc{txid}, the 4-byte position, and the 4-byte sequence
number; and a variable one, comprising the unlocking script and the encoding
of its size.

\subsection{Transaction output format}

Transaction outputs include two key pieces of information: an amount and a
locking script.

The amount of Bitcoin associated with the output is encoded using an 8-byte
integer. Like unlocking scripts, locking scripts have a variable size, which
is encoded using a varint. The contents of locking scripts depend on the type
of output that is created, and will be covered in the next section.

The output size is thus made up of a fixed component of eight bytes to specify
the amount, and a dynamic component comprising the unlocking script and the
encoding of its size.

\subsection{Transaction witness format}

Witnesses can serve as alternative stores for data to unlock outputs. Each
witness contains a varint to indicate the number of items it
contains. The items are of arbitrary size, so each item's length is encoded
using a varint as well.


\section{Transaction inputs, outputs, and witnesses}
\label{sec:analysis}

In the following, the inputs, outputs, and (where applicable) witnesses
formats of the different transaction types are investigated.  In each
instance, the discussion begins with a first-principles-based analysis of size
requirements; based on this, an estimate for the size requirements is derived,
which is then verified using empirical data (including historical data in
Bitcoin's blockchain up until November 2020).

\subsection{Pay-to-Public-Key}

\subsubsection{Outputs}

The locking-script format used by Pay-to-Public-Key (\textsc{p2pk}) outputs is
shown in Table~\ref{tab:locking}, with $\sigma_p$, the size of the following
public key; $p$, an \textsc{sec}-encoded public key; and
\textsc{op\_checksig}, the Bitcoin Script instruction for signature
verification. The encoding of the key's size requires one byte, the encoded
key uses 33~bytes, and the Script instruction uses another byte, leading to a
total script size of 35~bytes.

\begin{table}[!tb]
    \caption{Unlocking script formats by transaction type.}
    \begin{center}
        \begin{tabularx}{\linewidth}{|l|X|}
            \hline
            \textbf{Type}                           & \textbf{Unlocking script}   \\
            \hline
            \textsc{p2pk}                           & $\sigma_s$ $s$ \\
            \textsc{p2pkh}                          & $\sigma_s$ $s$ $\sigma_p$ $p$ \\
            Bare \textsc{ms}                        & \textsc{op\_0} $\sigma_{s_1}$ $s_1$ \ldots\,\, $\sigma_{s_m}$ $s_m$ \\
            \textsc{p2sh}-\textsc{ms}               & $d$ $\sigma_r$ $r$ \\
            \textsc{p2sh}-\textsc{p2wsh}-\textsc{ms}& $\sigma_r$ \textsc{op\_0} 32 $h_w$ \\
            \textsc{p2sh}-\textsc{p2wpkh}           & $\sigma_r$ \textsc{op\_0} 20 $h_p$ \\
            \hline
        \end{tabularx}
        \label{tab:unlocking}
    \end{center}
\end{table}

Together with the eight-byte amount and the one-byte varint to encode the
locking script's size, this leads to a total size of 44~bytes for
\textsc{p2pk} outputs.

\begin{figure}[tb]
    \centerline{\includegraphics[width=\linewidth]{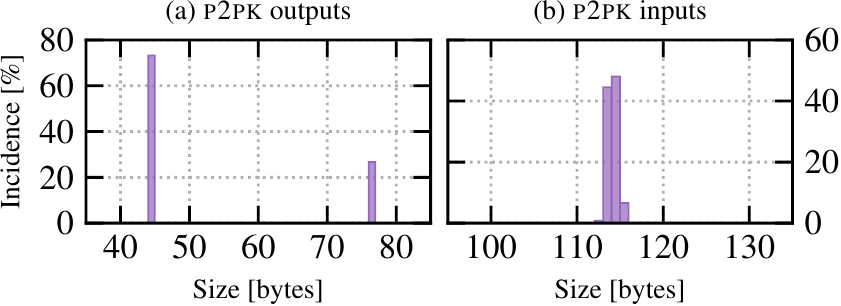}}
    \caption{Histograms of \textsc{p2pk} (a) output and (b) input sizes.}
    \label{fig:p2pk}
\end{figure}

This analytic estimate is validated by the empirical data shown in
Fig.~\ref{fig:p2pk}a, which contains a histogram of the sizes of \textsc{p2pk}
outputs. Discounting 76-byte outputs, which are an artifact
from Bitcoin's early days when public keys were encoded using the uncompressed
\textsc{sec} format, all outputs match the analytic estimate of 44~bytes.

\subsubsection{Inputs}

The unlocking-script format used by \textsc{p2pk} inputs is shown in
Table~\ref{tab:unlocking}, with $\sigma_s$, the size of the following
signature, and $s$, a \textsc{der}-encoded signature. The encoding of the size
of the signature requires one byte and the signature itself requires, on
average, 71.5~bytes, leading to an average total script size of 72.5~bytes.

Also taking into account the 40~bytes for \textsc{txid}, position, and
sequence number as well as one-byte varint to encode the script's size, this
leads to an average input size of 113.5~bytes.

As before, the analytic estimate is validated by empirical data.
Fig.~\ref{fig:p2pk}b contains a histogram of the sizes of all \textsc{p2pk}
inputs up to Nov.\ 2020. As expected, more than 90\% of all inputs have a size
of 113 or 114~bytes and thus match the estimate.  Inputs with a size of
115~bytes are artifacts from a time when the ``low $s$'' rule was not
enforced; inputs smaller than 113~bytes can be attributed to the occurrence of
encodings of $r$ and $s$ with eight or more leading zero bits (cf.\
Sect.~\ref{sec:encoding:sig}).


\subsection{Pay-to-Public-Key-Hash}

\subsubsection{Outputs}

Pay-to-Public-Key-Hash (\textsc{p2pkh}) outputs the locking-script
format shown in Table~\ref{tab:locking}, with
\textsc{op\_dup}, the Bitcoin Script instruction to duplicate the top stack
item; \textsc{op\_hash160}, the Bitcoin Script instruction to apply the
\textsc{hash}160 function\footnote{Bitcoin defines \textsc{hash}160 as
subsequently applying the \textsc{sha}256 and \textsc{ripemd}160 hashing
functions to input data.} to the top stack item; 20, the size of the
following hash; $h$, a 20-byte \textsc{hash}160 of a public
key; \textsc{op\_equalverify}, the Bitcoin Script instruction to make the
transaction invalid if the two top stack items differ; and
\textsc{op\_checksig}, the Bitcoin Script instruction to verify a signature.

The four Bitcoin Script instructions and the encoding of the size of the
hash require one byte each. Together with the 20-byte hash, this leads to
locking-script size of 25 bytes.

Taking into account the eight-byte amount and the one-byte varint to encode
the size of the locking script, the total size of \textsc{p2pkh} outputs is
34~bytes. As before, this analytic estimate is corroborated by empirical data,
shown in Fig.~\ref{fig:p2pkh}a. The data indicates that all \textsc{p2pkh}
outputs have a size of 34~bytes.

\subsubsection{Inputs}

\textsc{p2pkh} inputs use the unlocking-script format shown in
Table~\ref{tab:unlocking}, with $\sigma_s$, the size of the following
signature; $s$, a \textsc{der}-encoded signature; $\sigma_p$, the size of the
following public key; and $p$, a \textsc{sec}-encoded public key.

The encodings of the sizes of the signature and the public key require one
byte each, whereas signature and public key require 71.5 and 33 bytes,
respectively, leading to an average unlocking-script size of 106.5~bytes.

Considering the 40~bytes for \textsc{txid}, position, and sequence number as
well as a one-byte varint to encode the unlocking script's size, this results
in an average input size of 147.5~bytes.

\begin{figure}[tb]
    \centerline{\includegraphics[width=\linewidth]{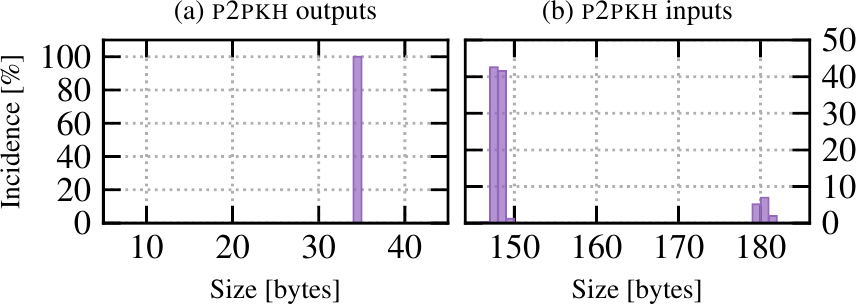}}
    \caption{Histograms of \textsc{p2pkh} (a) output and (b) input sizes.}
    \label{fig:p2pkh}
\end{figure}

The estimate is supported by empirical data shown in Fig.~\ref{fig:p2pkh}b.
As expected, the majority of inputs have a size of 147 or 148~bytes.  As
before, small deviations from the estimate can be attributed to the
\textsc{der} encoding. The second cluster around 180~bytes is an artifact
from Bitcoin's early days where public keys where encoded using the
uncompressed \textsc{sec} format.

\subsection{Bare Multi-Signature}
\label{sec:analysis:baremultisig}

\subsubsection{Outputs}

The (bare) multi-signature (\textsc{ms}) locking-script format is shown in
Table~\ref{tab:locking}, with \textsc{op\_}$m$, indicating the number of
signatures required to satisfy the locking script; $\sigma_{p_i}$ and $p_i$,
the sizes and encodings of $n$ public keys; \textsc{op\_}$n$, indicating the
number of public keys; and \textsc{op\_checkmultisig}, the
\textsc{ms}-validation Script instruction.

The Script instructions contribute three bytes, the encoding of the size of
each of the $n$ public keys contributes one byte, and each public key
33~bytes. This results in a locking-script size of $34n+3$~bytes.  Together
with the eight-byte amount and the one-byte varint to encode the script's
size, this results in a size of $34n+12$~bytes for $m$-of-$n$ \textsc{ms}
outputs.

\begin{figure}[b]
    \centerline{\includegraphics[width=\linewidth]{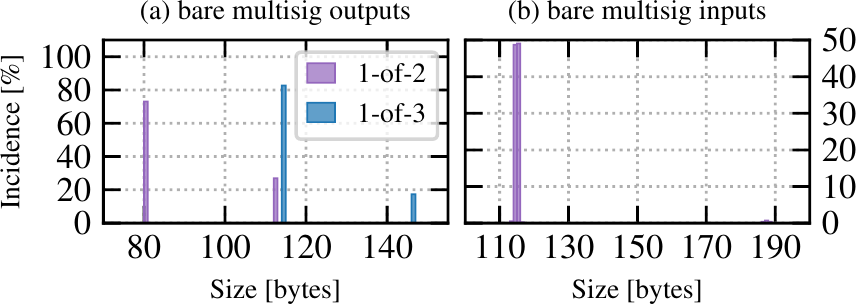}}
    \caption{Histograms of bare \textsc{ms} (a) output and (b) input sizes.}
    \label{fig:bare-multisig}
\end{figure}

This analytic estimate is verified for 1-of-2 and 1-of-3 \textsc{ms} outputs,
which together amount for more than 98\% of all \textsc{ms} outputs. The
estimates are 80 and 114~bytes, respectively. The empirical data shown in
Fig.~\ref{fig:bare-multisig}a supports these estimates: the bulk of 1-of-2 and
1-of-3 \textsc{ms} outputs have a size of 80 and~114 bytes, respectively. In
each case, there is a smaller number of outputs that are 32~bytes larger than
the estimate---artifacts from old transaction using the uncompressed format
in which public key's include the 32-byte $y$-coordinate.

\subsubsection{Inputs}

The \textsc{ms} unlocking-script format is shown in Table~\ref{tab:unlocking},
with \textsc{op\_0}, a dummy Bitcoin Script instruction to address a bug in
the implementation of \textsc{op\_checkmultisig}; and $\sigma_{s_i}$ and
$s_i$, the sizes and encodings of $m$ signatures.

The Script instruction contributes one byte, the encodings of the size of the
$m$ signatures one byte each, and each signature, on average, requires
71.5~bytes. This results in an average unlocking-script size of
$72.5m+1$~bytes.  Also considering the 40~bytes for \textsc{txid}, position,
and sequence number and the one-byte varint to encode the script's size, the
resulting average size for $m$-of-$n$ \textsc{ms} inputs is $72.5m+42$~bytes.

The analytic estimate is verified for 1-of-2 and 1-of-3 \textsc{ms} variants,
which together amount for more than 98\% of all \textsc{ms} inputs. Since
$m$=1 for both variants, they share the estimate of 114.5~bytes, which is
supported by the empirical data shown in Fig.~\ref{fig:bare-multisig}b: as
expected, the bulk of all 1-of-2 and 1-of-3 \textsc{ms} outputs have a size of
either 114 or 115~bytes.


\subsection{Null Data}

\subsubsection{Outputs}

The Null-Data locking-script format is shown in Table~\ref{tab:locking}, with
\textsc{op\_return}, the Bitcoin Script instruction to indicate an unspendable
output; $\sigma_d$, the size of the following data; and $d$, the data included
in the output. Note that as of Bitcoin Core 0.12.0, only 80~bytes of data are
allowed.

The Bitcoin Script instruction contributes one byte. The encoding of the
data's size requires one or two bytes;\footnote{\label{fn:constantsize}In
Bitcoin Script, sizes between 1--75~bytes are implicitly encoded using a
single byte; larger values require a one-byte magic number to indicate an
explicitly encoded length in one, two, or four additional
bytes~\cite{wiki:script:constants}.} and the actual data requires $\sigma_d$
bytes.  The locking-script size is thus $\sigma_d+2$ for data smaller than
75~bytes and $\sigma_d+3$ for larger data. Combined with the eight-byte amount
and the one-byte varint to encode the script's size, this results in an output
size of $\sigma_d+11$ for outputs that include up to 75~bytes of data, and
$\sigma_d+12$ for outputs with more data.

\begin{figure}[tb]
    \centerline{\includegraphics{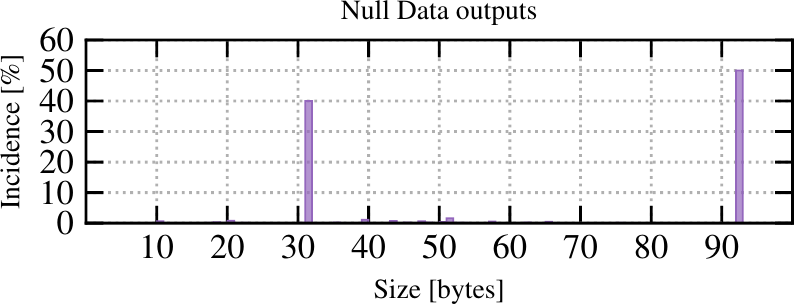}}
    \caption{Histogram of Null Data (\texttt{OP\_RETURN}) output sizes.}
    \label{fig:nulldata}
\end{figure}

This analytic estimate is verified for 20- and 80-byte Null-Data outputs,
which together amount for more than 90\% of all Null-Data outputs. The
estimate for the former is 31~bytes; for the latter it is 92 bytes. These
estimates are corroborated by the empirical data in Fig.~\ref{fig:nulldata}.

\subsection{Pay-to-Script-Hash}

\subsubsection{Outputs}

The Pay-to-Script-Hash (\textsc{p2sh}) locking-script format is shown in
Table~\ref{tab:locking}, with \textsc{op\_hash160}, the Bitcoin Script
instruction to apply the \textsc{hash}160 function; 20, the size of the
following hash; $h_r$, a 20-byte \textsc{hash}160 of a redeem script;
and \textsc{op\_equal}, the Script instruction to determine whether the two
top stack items are identical.

The Script instructions and the encoding of the size of the hash contribute
one byte each; together with the 20-byte hash, this results in a total locking
script size of 23~bytes. Together with the eight-byte amount and the one-byte
varint encoding the script's size, this leads to an output size of 32~bytes.
This estimate is confirmed by the empirical data shown in
Fig.~\ref{fig:P2SH-output-multisig-input}a.

In contrast to fixed-size \textsc{p2sh} outputs, the size of \textsc{p2sh}
inputs varies significantly depending on the type of redeem script included in
the input. The most relevant use cases are discussed in the
following.


\subsubsection{\textsc{p2sh}-\textsc{ms} inputs}

The \textsc{p2sh}-\textsc{ms} unlocking-scripts format is shown in
Table~\ref{tab:unlocking}, with $d$, data corresponding to the signatures
required to satisfy the redeem script; $\sigma_r$, the size of the following
redeem script; and $r$, the redeem script.  In case of
\textsc{p2sh}-\textsc{ms}, the data, $d$, follows the conventions for
\textsc{ms} unlocking scripts documented in Table~\ref{tab:unlocking}, whereas
the redeem script, $r$, follows those of \textsc{ms} locking scripts
documented in Table~\ref{tab:locking}.  In
Sect.~\ref{sec:analysis:baremultisig}, the sizes for these scripts were
established to be $72.5m+1$ and $34n+3$~bytes, respectively.  In case the
redeem script is smaller than 76~bytes, the encoding of its size, $\sigma_r$,
requires one byte; in case it is larger, two
bytes.\footnotemark[6] The unlocking scripts used in
$m$-of-$n$-\textsc{p2sh}-\textsc{ms} inputs thus have a size of
$72.5m+34n+5$~bytes for redeem scripts smaller than 76~bytes, and an extra
byte in case of larger redeem scripts.

Together with the 41-byte contribution of \textsc{txid}, position, sequence
number, and the encoding of the script's size, this leads to an estimate of
$72.5m+34n+46$ for redeem scripts smaller than 76~bytes; larger redeem scripts
are subject to additional overhead (discussed in more detail in the
following).

These analytic estimates are verified for 2-of-2 and 2-of-3
\textsc{p2sh}-\textsc{ms} inputs, which together amount for more than 90\% of
such inputs. For the former, $n=2$, so the estimate for the redeem script's
size is $34n+3=71$~bytes. The encoding of the script's size therefore requires
only one byte. For $m=n=2$, the estimate of the input's size is thus
$72.5m+34n+46=259$~bytes. For 2-of-3 \textsc{p2sh}-\textsc{ms} inputs, $n=3$,
and the average redeem-script size is $34n+3=105$~bytes, which means the
encodings of the redeem script's size requires two
bytes.\footnotemark[6] Together with the data to satisfy
the redeem script, which requires $72.5m+1=146$~bytes, this leads to a total
unlocking-script size of $105+2+146=253$. Because varints can only encode
numbers up to 252 with a single byte, the encoding of the unlocking script's
size requires three bytes. Taking into account the contributions of the
32-byte \textsc{txid}, the 4-byte position, the 4-byte sequence number, the
3-byte varint to encode the script's size, and the 253-byte unlocking script,
yields a total input size of 296~bytes.

\begin{figure}[tb]
    \centerline{\includegraphics[width=\linewidth]{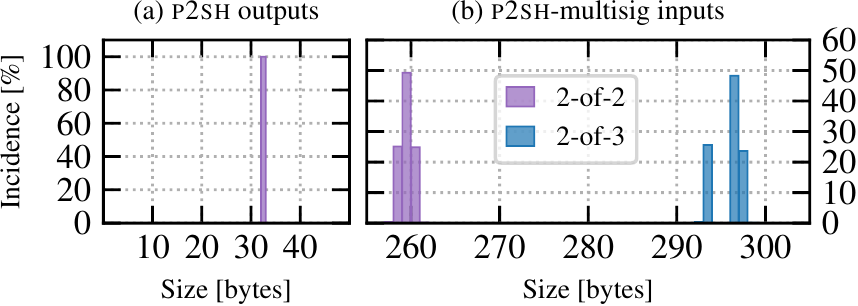}}
    \caption{Histograms of \textsc{p2sh} (a) output and (b) \textsc{p2sh}-\textsc{ms} input sizes.}
    \label{fig:P2SH-output-multisig-input}
\end{figure}

The two estimates are corroborated by empirical data in
Fig.~\ref{fig:P2SH-output-multisig-input}b.  Almost half of all 2-of-2
\textsc{p2sh}-\textsc{ms} inputs match the estimate of 259 bytes; signatures one
byte smaller or larger than the estimate are explained by \textsc{der}-encoded
signatures being, on average, 71.5 bytes: absent the ``low $r$'' optimization,
there is a 50\% probability of generating either a 71-byte or 72-byte
signature; for two signatures, this implies a 25\% chance of two 71-byte
signatures (corresponding to inputs with a size of 258 bytes), a 50\% chance
of one 71-byte and a 72-byte signature (resulting in 259-byte inputs), and a
25\% chance of two 72-byte signatures (leading to 260-byte inputs).

For 2-of-3 \textsc{p2sh}-\textsc{ms}, too, almost half of all inputs match the
estimate of 296~bytes.  Again, around 25\% of inputs are one byte
larger than the estimate, which, as before, can be explained by a 25\% chance
of generating two 72-byte signatures. Another 25\% of inputs have a size of
293~bytes, three bytes smaller than the estimate, which can be explained by a
25\% chance of generating two 71-byte signatures.  In this instance, the
unlocking script's size is only 252, which means the script's size can be
encoded using a one-byte varint instead of a three-byte one. The three byte
deviation from the estimate thus originates from: a one-byte reduction caused
by the use of two 71-byte signatures instead of the estimate of two times 71.5
bytes; and two more bytes because only one byte (instead of three) is required
to encode the unlocking script's size.


\subsubsection{\textsc{p2sh}-Pay-to-Witness-Script-Hash-\textsc{ms} Inputs}


The unlocking-script format of
\textsc{p2sh}-Pay-to-Witness-Script-Hash-\textsc{ms}
(\textsc{p2sh}-\textsc{p2wsh}-\textsc{ms}) inputs is documented in
Table~\ref{tab:unlocking}, with $\sigma_r$, the size of the following
redeem script; and the redeem script, comprising \textsc{op\_0}, the Script
instruction to indicate a version zero witness program; 32, the size of
the following hash; and $h_w$, a 32-byte \textsc{sha}256 hash of the
witness script.

\begin{figure}[tb]
    \centerline{\includegraphics[width=\linewidth]{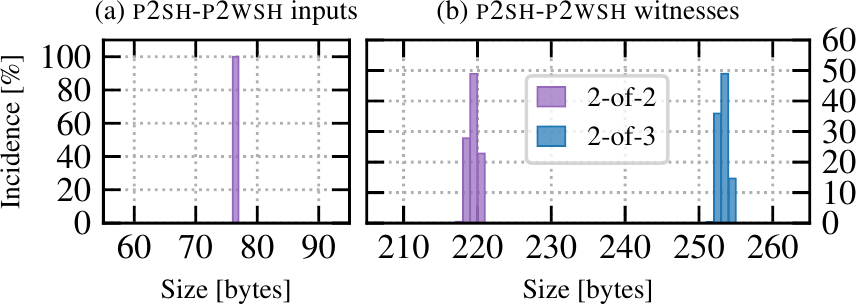}}
    \caption{Histograms of \textsc{p2sh}-\textsc{p2wsh} (a) input and (b) witness sizes.}
    \label{fig:P2SH-P2WSH}
\end{figure}

The encoding of the redeem script's size, the Script instruction, and the
encoding of the hash's size contribute one byte each. Together with 32-byte
witness script hash, this results in an unlocking-script size of 35~bytes.
Taking into account the contribution of the 32-byte \textsc{txid}, the 4-byte
position, the 4-byte sequence number, and one byte for the varint to encode
the unlocking script's size, this results in a total input size of 76~bytes.
This estimate is validated by the empirical data shown in
Fig.~\ref{fig:P2SH-P2WSH}a, which confirms all \textsc{p2sh}-\textsc{p2wsh}
inputs have a size of 76~bytes.

\subsubsection{\textsc{p2sh}-\textsc{p2wsh}-\textsc{ms} Witnesses}
\label{sec:analysis:p2sh-p2wsh:witnesses}

The \textsc{p2sh}-\textsc{p2wsh} witness format is documented in
Table~\ref{tab:witness}, with $n_\mathrm{i}$, a varint encoding the number
of items in the witness; $d$, data, consisting of $m$ signatures (following
the format of \textsc{ms} unlocking scripts shown in Table~\ref{tab:unlocking}) to
satisfy the witness script; $\sigma_w$, the size of the witness script;
and $w$, a witness script, containing the parameters $m$ and $n$, as well as
$n$ public keys (following the format of \textsc{ms} locking scripts documented
in Table~\ref{tab:locking}), that will be interpreted as locking script.

\begin{table}[b]
    \caption{Witness formats by transaction type.}
    \begin{center}
        \begin{tabularx}{\linewidth}{|l|X|}
            \hline
            \textbf{Type}                               & \textbf{Witness}   \\
            \hline
            \textsc{p2sh}-\textsc{p2wsh}-\textsc{ms}    & $n_\mathrm{i}$ $d$ $\sigma_w$ $w$ \\
            \textsc{p2sh}-\textsc{p2wpkh}               & 2 $\sigma_s$ $s$ $\sigma_p$ $p$ \\
            \textsc{p2wpkh}                             & 2 $\sigma_s$ $s$ $\sigma_p$ $p$ \\
            \textsc{p2wsh}-\textsc{ms}                  & $n_\mathrm{i}$ $d$ $\sigma_w$ $w$ \\
            \textsc{p2tr} (key path)                    & 1 $\sigma_s$ $s$  \\
            \textsc{p2tr} (script path)                 & $n_\mathrm{i}$ $d$ $\sigma_w$ $w$ $c$ \\
            \hline
        \end{tabularx}
        \label{tab:witness}
    \end{center}
\end{table}

Note that the number of witness items and the size of the witness script are
encoded using varints, not Bitcoin script instructions. In most cases, there
are less than 252 items in the witness, and the witness script is smaller than
252 bytes, so the varints encoding these typically contribute one byte each.
The contribution of the data to satisfy the unlocking script corresponds to
the estimate for \textsc{ms} locking scripts of $34n+3$~bytes established in
Sect.~\ref{sec:analysis:baremultisig}.\footnote{The fact that the public keys'
sizes are encoded with a one-byte varint instead of a one-byte Bitcoin Script
instruction does not affect the estimate.}
The witness script's contribution corresponds to the estimate for \textsc{ms}
unlocking scripts of $72.5m+1$ bytes (cf.\
Sect.~\ref{sec:analysis:baremultisig}). So, in case there are fewer than
252~items in the witness and the witness script is at most 252~bytes, the
overall estimate for the witness is $72.5m+34n+6$~bytes.

The analytic estimate is verified for 2-of-2 and 2-of-3
\textsc{p2sh}-\textsc{p2wsh}-\textsc{ms} inputs, which together amount for
more than 90\% of such inputs. For the former, $m=n=2$, so the estimate is
$72.5m+34n+6=219$~bytes; for the latter, $m=3$ and $n=2$, and the estimate is
253~bytes.

The estimates are corroborated by the empirical data shown in
Fig.~\ref{fig:P2SH-P2WSH}b, which contains a histogram of all 2-of-2 and 2-of-3
\textsc{p2sh}-\textsc{p2wsh}-\textsc{ms} witnesses up to Nov.\ 2020. For both
\textsc{ms} variants, the peak of its distribution matches the estimate. As before,
inputs that are one byte smaller or larger than the estimate can be explained by
the 25\% chances of creating either two 71-byte signatures or two 72-byte
signatures instead of the estimate, which uses the average of two times
71.5~bytes.


\subsubsection{\textsc{p2sh}-Pay-to-Witness-Public-Key-Hash Inputs}
\label{sec:analysis:p2sh-p2wpkh-inputs}

The unlocking-script format of \textsc{p2sh}-Pay-to-Witness-Public-Key-Hash
(\textsc{p2sh}-\textsc{p2wpkh}) inputs is documented in
Table~\ref{tab:unlocking}, with $\sigma_r$, the size of the following
redeem script; and the redeem script, comprising \textsc{op\_0}, the Script
instruction to indicate a version zero witness program; 20, the size of
the following hash; and $h_w$, a 20-byte \textsc{hash}160 hash of a
public key.

\begin{figure}[tb]
    \centerline{\includegraphics[width=\linewidth]{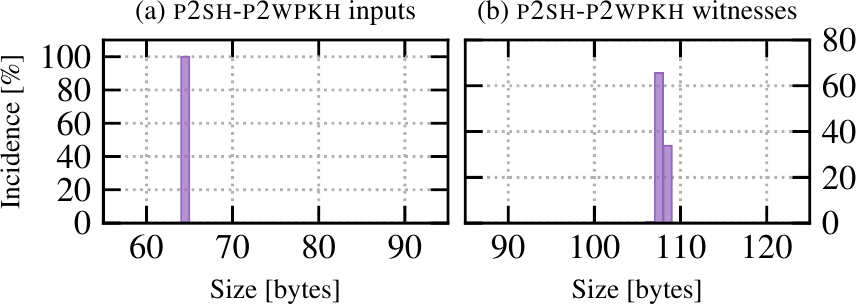}}
    \caption{Histograms of \textsc{p2sh}-\textsc{p2wpkh} (a) input and (b) witness sizes.}
    \label{fig:P2SH-P2WPKH}
\end{figure}

The encoding of the sizes of the redeem script and the hash of the public key
contribute one byte each.  Together with the one-byte Script instruction and
the 20-byte hash, this results in a total unlocking script size of 23~bytes.
Together with the 32-byte \textsc{txid}, the 4-byte position, the 4-byte
sequence number, and one byte for the varint to encode the unlocking script's
size, this results in a total input size of 64~bytes.  This estimate is
validated by the empirical data in Fig.~\ref{fig:P2SH-P2WPKH}a, which confirms
that all \textsc{p2sh}-\textsc{p2wpkh} inputs have a size of 64~bytes.

\subsubsection{\textsc{p2sh}-\textsc{p2wpkh} Witnesses}
\label{sec:analysis:p2sh-p2wpkh-witnesses}

The \textsc{p2sh}-\textsc{p2wpkh} witness format is documented in
Table~\ref{tab:witness}, with 2, to indicate two witness items;
$\sigma_s$, a varint encoding the size of the following signature;
$s$, a valid signature for the following public key; $\sigma_p$, a
varint encoding the size of the following public key; and $p$, the public key
corresponding to the \textsc{hash}160 hash used in the locking script.
The three varints contribute one byte each, and the signature and key 71.5 and
33~bytes, respectively, making for a total witness size of 107.5~bytes.

This estimate is corroborated by empirical data shown in
Fig.~\ref{fig:P2SH-P2WPKH}b, which indicates that more than 99\% of all
\textsc{p2sh}-\textsc{p2wpkh} witnesses have a size of either 107 or
108~bytes.  The bias toward 107~bytes can be explained by the fact that the
``low $r$'' optimization, which results in 71-byte signatures in comparison to
the 71.5-byte average used by the estimate, was already widely used by the
time \textsc{p2sh}-\textsc{p2wpkh} was introduced.


\subsection{Pay-to-Witness-Public-Key-Hash}

\subsubsection{Outputs}

The locking-script format of Pay-to-Witness-Public-Key-Hash (\textsc{p2wpkh})
outputs is shown in Table~\ref{tab:locking}, with \textsc{op\_0}, the Script
instruction to indicate a version zero witness program; 20, the size of the
following hash; and $h_p$, a \textsc{hash}160 hash of a public key. The
Bitcoin script instruction and the encoding of the hash's size
contribute one byte each, the last 20~bytes, resulting in a
locking-script size of 22~bytes.

\begin{figure}[tb]
    \centerline{\includegraphics[width=\linewidth]{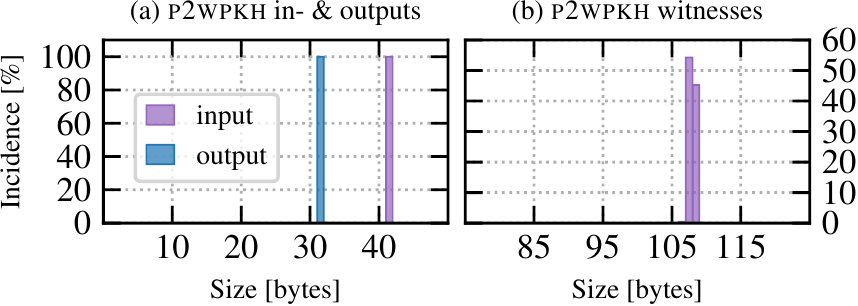}}
    \caption{Histograms of \textsc{p2wpkh} (a) input and output, and (b) witness sizes.} 
    \label{fig:P2WPKH}
\end{figure}

Together with the 8-byte amount and 1-byte varint to encode the script's
size, this results in an output size of 31~bytes.  This estimate is
validated by the empirical data in Fig.~\ref{fig:P2WPKH}a, which confirms that
all \textsc{p2wpkh} outputs have a size of 31~bytes.

\subsubsection{Inputs}
\label{sec:analysis:P2WPKH:inputs}

For \textsc{p2wpkh}, the data to satisfy the locking script resides in the
witness, so the unlocking script is empty. \textsc{p2wpkh} inputs thus consist
only of a 32-byte \textsc{txid}, a 4-byte position, a 4-byte sequence number,
and a 1-byte varint to indicate a zero-length unlocking script. Inputs thus
have a fixed size of 41~bytes, a fact which is corroborated by empirical data
shown in Fig.~\ref{fig:P2WPKH}a.


\subsubsection{Witnesses}
\label{sec:analysis:P2WPKH:witnesses}

\textsc{p2wpkh} witnesses contain the same data as
\textsc{p2sh}-\textsc{p2wpkh} witnesses (cf.\ Table~\ref{tab:witness}): 2, to
indicate two witness items; $\sigma_s$ and $s$, the size of and the corresponding
signature; and $\sigma_p$ and $p$, the size of and the corresponding public key.
Again, the varints contribute one byte each, the signature and public key 71.5
and 33~bytes, respectively, resulting in a witness-size estimate of
107.5~bytes.

The estimate is validated by empirical data shown in Fig.~\ref{fig:P2WPKH}b,
which shows that 99\% of all \textsc{p2wpkh} witnesses have a size of either
107 or 108~bytes.  As was the case for \textsc{p2sh}-\textsc{p2wpkh}
witnesses, the fact that SegWit was introduced after the ``low $r$''
optimization explains the bias toward 107-byte witnesses (cf.\
Sect.~\ref{sec:analysis:p2sh-p2wpkh-witnesses}).



\subsection{Pay-to-Witness-Script-Hash}

\subsubsection{Outputs}

The Pay-to-Witness-Script-Hash (\textsc{p2wsh}) locking-script format is shown
in Table~\ref{tab:locking}, with \textsc{op\_0}, the Script instruction to
indicate a version zero witness program; 32, the size of the following hash;
and $h_w$, the \textsc{sha}256 hash of a witness script. The first two items
contribute one byte each, the last 32~bytes, resulting in a locking-script
size of 34~bytes.

Also considering the account the amount (eight~bytes) and the varint encoding
the script's size (one byte), this results in an output size of 43~bytes.
This estimate is validated by the empirical data in Fig.~\ref{fig:P2WSH}a,
which confirms that all \textsc{p2wsh} outputs have a size of 43~bytes.


\subsubsection{Inputs}

For \textsc{p2wsh}, the data to satisfy the locking script resides in the
witness, so the unlocking script is empty, as was the case for
\textsc{p2wpkh}. Inputs thus have a fixed size of 41~bytes (cf.\
Sect.~\ref{sec:analysis:P2WPKH:inputs}). This estimate is corroborated by
empirical data shown in Fig.~\ref{fig:P2WSH}a.


\subsubsection{Witnesses}

More than 98\% of all \textsc{p2wsh} transactions are used for \textsc{ms}, so
the following discussion focuses on \textsc{p2wsh}-\textsc{ms}.
\textsc{p2wsh}-\textsc{ms} witnesses contain the same data as
\textsc{p2sh}-\textsc{p2wsh}-\textsc{ms} witnesses (cf.\
Table~\ref{tab:witness}).  The estimate for the witness size is thus identical
as well and corresponds to $72.5m+34n+6$~bytes (cf.\
Sect.~\ref{sec:analysis:p2sh-p2wsh:witnesses}).

In the following, this analytic estimate is verified for 1-of-1, 2-of-2, and
2-of-3 \textsc{p2wsh}-\textsc{ms} witnesses, which together amount for more than
98\% of all \textsc{p2wsh}-\textsc{ms} transactions.

\begin{figure}[tb]
    \centerline{\includegraphics[width=\linewidth]{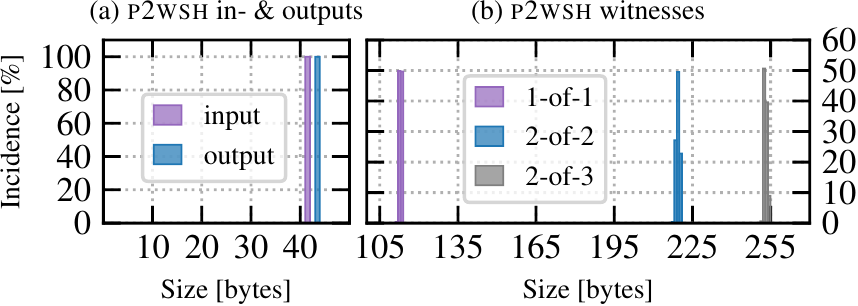}}
    \caption{Histograms of \textsc{p2wsh} (a) input and output, and (b) witness sizes.}
    \label{fig:P2WSH}
\end{figure}

For $m=n=1$, the estimate is 112.5~bytes; for $m=n=2$, it is 219 bytes; and for
$m=2$ and $n=3$, 253~bytes. All estimates are supported by the empirical
data shown in Fig.~\ref{fig:P2WSH}b. In all instances, the observed sizes
match the analytic estimates. For the latter two variants, the bias toward
smaller sizes can again be explained by the ``low $r$'' optimization (cf.\
Sections~\ref{sec:analysis:p2sh-p2wpkh-witnesses} and~\ref{sec:analysis:P2WPKH:witnesses}).


\subsection{Pay-to-Taproot}
\label{sec:analysis:P2TR}

Since Pay-to-Taproot (\textsc{p2tr}) is not yet available,
estimates in this section cannot be validated using empirical data.

\subsubsection{Outputs}
\label{sec:analysis:P2TR:outputs}

The \textsc{p2tr} locking-script format is shown in Table~\ref{tab:locking},
with \textsc{op\_1}, the Script instruction to indicate a version-one witness
program; 32, the size of the following hash; and $p$, a 32-byte tweaked
Schnorr public key. The first two items each contribute one, the last
32~bytes, resulting in a locking-script size of 34~bytes. Adding the 8-byte amount
and one-byte varint to encode the script's size, yields an output size
of 43~bytes.


\subsubsection{Inputs}
\label{sec:analysis:P2TR:inputs}

The data to satisfy the locking script resides in the witness, leaving the
unlocking script empty. As in case of \textsc{p2wpkh} (cf.\
Sect.~\ref{sec:analysis:P2WPKH:inputs}), inputs thus have a size of 41~bytes.


\subsubsection{Witnesses}
\label{sec:analysis:P2TR:witnesses}

\textsc{p2tr} locking scripts can be satisfied either by \emph{key
path} (i.e., by providing a valid signature for the tweaked public key in the
locking script) or \emph{script path} (i.e., by providing a valid input,
a corresponding witness script, an untweaked public key, and hashes of the
leaves and branches of the Merkle tree required to determine the tree's root).

In case of the key path, \textsc{p2tr} witnesses comprise: 2, the number of
witness items; $\sigma_s$, the size of the following Schnorr signature; and
$s$, a Schnorr signature. The two first items are varints that contribute one
byte each, and a Schnorr signature uses 64~bytes in case of the default
signature hash type and 65~bytes in case of a custom signature hash type.
Absent empirical data, the witness-size estimate will be based on the
assumption that the default signature hash type is used in most cases. The
\textsc{p2tr} key-path witness-size estimate based on this hypothesis is,
therefore, 66~bytes.

In case of the script path, \textsc{p2tr} witnesses comprise: $n_\mathrm{i}$,
the number of witness items; $d$, data to satisfy the script presented next;
$\sigma_w$, the size of the following witness script; $w$ a witness script to
be interpreted as locking script; and $c$, a control block (the first byte of
the control block encodes the leaf version, which in case of the script path
is always 0xc0; the next 32 bytes encode an untweaked Schnorr public key;
finally, the control block holds one or more 32-byte blocks that encode the
hashes of the leaves and branches of the Merkle tree that are necessary to
reconstruct the Merkle root used to tweak the public key).




\section{Automating transaction-size estimates}
\label{sec:libtxsize}

\emph{libtxsize} integrates the previously established and empirically
validated analytic models to estimate the size, virtual size and weight of
arbitrary transactions.  In addition to overall estimates, \emph{libtxsize}
can provide information about a transaction's components, such as the sizes of
individual inputs, outputs, and witnesses, as well as transaction overhead.

The library is available on
GitHub\footnote{\texttt{https://github.com/virtu/libtxsize}} and uses a
bottom-up approach to create estimates: first, if necessary, the sizes of
redeem and witness scripts are estimated to determine the size requirements of
the Bitcoin Script instructions and varints that encode the lengths of such
scripts; next, the script and witness sizes are calculated, taking the results
of the previous step into account. Once the script and witness sizes are
known, the size of the Bitcoin Script instructions and varints to encode their
length are determined, and other constant contributions of inputs
(\textsc{txid}, position, and sequence number) and outputs (amount) are
considered. Finally, the sizes of all inputs, outputs, and witnesses are
assembled, and the transaction overhead (transaction version, two varints
indicating the number of inputs and outputs, lock time, and, if applicable,
SegWit marker and version) is added.

The library is written in Python and exposes a Python interfaces to get
estimates for input, output, and witnesses sizes, as well as estimates for
transactions. \emph{libtxsize} also includes a command-line interface to
facilitate quick experimentation.


\section{Results and Conclusion}
\label{sec:results}

The formats of Bitcoin transactions, inputs, outputs, and witnesses were
presented. Moreover, the sizes of different input, output, and witness types
were investigated using first-principles analysis, from which analytic
estimates were derived. A summary of these estimates is presented in
Table~\ref{tab:estimates}. Furthermore, all estimates (with the exception of
Pay-to-Taproot, for which no empirical data is available so far) were
validated using empirical data.

\begin{table}[!tb]
    \caption{Estimates of components (in bytes) by transaction type.}
    \begin{center}
        \begin{threeparttable}
            \resizebox{\linewidth}{!}{
            \begin{tabular}{|l|r|r|r|}
                \hline
                \textbf{Type}                                       & \textbf{Output\,[B]}  & \textbf{Input\,[B]}                               & \textbf{Witness\,[B]}     \\
                \hline
                \textsc{p2pk}                                       & 44                    & 113.5                                             & ---                       \\
                \textsc{p2pkh}                                      & 34                    & 147.5                                             & ---                       \\
                \textsc{p2wpkh}                                     & 31                    & 41                                                & 107.5                     \\
                \textsc{p2tr} (key path)                            & 43                    & 41                                                & 66                        \\
                \textsc{p2wsh}-$m$-of-$n$-\textsc{ms}               & 43                    & 41                                                & 72.5$m$+34$n$+6\tnote{\textsection}           \\
                \textsc{p2sh}-$m$-of-$n$-\textsc{ms}                & 32                    & 72.5$m$+34$n$+46\tnote{\textasteriskcentered}     & ---                       \\
                \textsc{p2sh}-\textsc{p2wsh}-$m$-of-$n$-\textsc{ms} & 32                    & 76                                                & 72.5$m$+34$n$+6\tnote{\textsection}           \\
                \textsc{p2sh}-\textsc{p2wpkh}                       & 32                    & 64                                                & 107.5                     \\
                \textsc{null data}                                  & $\sigma_d$+11\tnote{\dag}    & ---                                               & ---                       \\
                Bare $m$-of-$n$-\textsc{ms}                         & 34$n$+12              & 72.5$m$+42                                        & ---                       \\
                \hline
            \end{tabular}
            }
            \label{tab:estimates}
            \begin{tablenotes}
                \item[\textsection]{Estimate two bytes larger if witness contains more than 252~items;\\
                    two more bytes required if witness script larger than 252~bytes.}
                \item[\textasteriskcentered]{Estimate one byte larger if redeem script larger than 75~bytes;\\
                    two more bytes required if unlocking script larger than 252~bytes.}
                \item[\dag]{If $\sigma_d>75$~bytes, the estimate is one byte larger.}
            \end{tablenotes}
        \end{threeparttable}
    \end{center}
\end{table}

Finally, \emph{libtxsize}, a library that makes the findings easily accessible
by automating estimates for the size, virtual size, and weight of transactions
and their components, was presented.



\bibliographystyle{IEEEtran}
\bibliography{IEEEabrv,citations}


%
%

\end{document}